\documentclass[amsmath,amssymb,preprint,10pt]{iopart}
\expandafter\let\csname equation*\endcsname\relax
\expandafter\let\csname endequation*\endcsname\relax 
\usepackage{amsmath}
\usepackage{amssymb}
\usepackage{graphicx}
\usepackage{dcolumn}
\usepackage{bm}
\usepackage[mathlines]{lineno}

\usepackage[utf8]{inputenc}
\usepackage[T1]{fontenc}
\usepackage{mathptmx}
\usepackage{subfig}
\usepackage[inline]{enumitem}
\usepackage{caption}
\usepackage{etoolbox}
\usepackage{color}
\usepackage{float}
\usepackage{xcolor}

\begin{document}

\title[]{Anomalous hot electron generation from two-plasmon decay instability driven by broadband laser pulses with intensity modulations}

\author{C. Yao$^{1,2}$, J. Li$^{1,3,a)}$, L. Hao$^{4,b)}$, R. Yan$^{5,3,c)}$, C. Wang$^6$, A. Lei$^{7,6}$, Y-K. Ding$^4$, J. Zheng$^{1,3}$}

\address{$^1$Department of Plasma Physics and Fusion Engineering and CAS Key Laboratory of Geospace Environment, University of Science and Technology of China, Hefei, Anhui 230026, People’s Republic of China}
\address{$^2$Research Center of Laser Fusion, Mianyang, Sichuan 621900, People’s Republic of China.}
\address{
$^3$Collaborative Innovation Center of IFSA (CICIFSA), Shanghai Jiao Tong University, Shanghai 200240, People’s Republic of China
}%
\address{%
$^4$Institute of Applied Physics and Computational Mathematics, Beijing 100088, People’s Republic of
China
}%
\address{ 
$^5$Department of Modern Mechanics, University of Science and Technology of China, Hefei, Anhui
230026, People’s Republic of China
}%
\address{
$^6$Shanghai Institute of Laser Plasma, China Academy of Engineering Physics, Shanghai 201800, People’s
Republic of China
}%
\address{
$^7$Joint Laboratory on High Power Laser and Physics, Shanghai Institute of Optics and Fine Mechanics, Chinese Academy of Sciences, Shanghai 201800, People’s Republic of China
}%

\ead{\mailto{a)junlisu@ustc.edu.cn}; \mailto{b)hao\_liang@iapcm.ac.cn}; \mailto{c)ruiyan@ustc.edu.cn}}


\begin{abstract}
We investigate the hot electrons generated from two-plasmon decay (TPD) instability driven by laser pulses with intensity modulated by a frequency $\Delta \omega_m$.
Our primary focus lies on scenarios where $\Delta \omega_m$ is on the same order of the TPD growth rate $ \gamma_0$ ( $\Delta \omega_m \sim \gamma_0$), 
corresponding to moderate laser frequency bandwidths for TPD mitigation. 
With $\Delta \omega_m$ conveniently modeled by a basic two-color scheme of the laser wave fields in fully-kinetic particle-in-cell simulations, we demonstrate that the energies of TPD modes and hot electrons exhibit intermittent evolution at the frequency $\Delta \omega_m$, particularly when $\Delta \omega_m \sim \gamma_0$. With the dynamic TPD behavior, the overall ratio of hot electron energy to the incident laser energy, $f_{hot}$, changes significantly with $\Delta \omega_m$. While $f_{hot}$ drops notably with increasing $\Delta \omega_m$ at large $\Delta \omega_m$ limit as expected, it goes anomalously beyond the hot electron energy ratio for a single-frequency incident laser pulse with the same average intensity when $\Delta \omega_m$ falls below a specific threshold frequency $\Delta \omega_c$. We find this threshold frequency primarily depends on $\gamma_0$ and the collisional damping rate of plasma waves, with relatively lower sensitivity to the density scale length. We develop a scaling model characterizing the relation of $\Delta \omega_c$ and laser plasma conditions, 
enabling the potential extention of our findings to more complex and realistic scenarios.
\end{abstract}

\maketitle
%
%
%
%
\ioptwocol

\section{introduction}\label{sec:level}
Two-plasmon decay (TPD) instability\cite{liu1976,afeyan1997a,Simon1983} is one of the primary laser plasma instabilities\cite{Kruer2003} (LPI) of inertial confinement fusion\cite{Atzeni2004,craxton2015} (ICF), in which a spherical target capsule is imploded by multiple overlapping laser beams to achieve ignition conditions. In both direct or indirect drive ICF experiments\cite{Seka2009a,Regan2010}, TPD generates hot electrons that can degrade the implosion by preheating the inner cold fuels before compression\cite{Kirkwood2013a,baton2020,christopherson2021,christopherson2022}, or anomalously absorbing\cite{turnbull2020} the pump laser, leading to a non-negligible laser energy loss. Generally, mitigating TPD is crucial for ICF implosion.

As a three-wave coupling LPI process, TPD growth relies on the frequency and wavenumber matching conditions\cite{Kruer2003}

\begin{align}
\label{eqn:w_match}
\omega_0&=\omega_1+\omega_2 \nonumber \\
\vec{k}_0&=\vec{k}_1+\vec{k}_2
\end{align}
where $\omega$ and $\vec{k}$ are the frequency and wavenumber with the subscripts 0, 1 and 2 denoting the pump laser wave and two daughter electron plasma waves (EPW), respectively. 
The matching conditions show that the pumps with different $\omega_0$'s should not drive the same TPD modes, so 
the broadband lasers with energy deployed in a considerable frequency range $\Delta\omega$ should inhibit the TPD growth, as the power for each $\omega_0$ component becomes weaker. 
Besides, employing mitigation methods on laser bandwidth may allow for more space of pulse shape and target designs\cite{Follett2016} to achieve higher hydrodynamic efficiency. Furthermore, recent experiments have shown that the bandwidth introduced by SSD smoothing\cite{Skupsky1989} can improve the low adiabat implosions considerably. All these suggest the great potential of broadband lasers in suppressing LPI and achieving high gain ICF.

Earlier research\cite{thomson1974,JJThomson1975} has predicted the inhibition of linear growth in LPI when the laser bandwidth $\Delta \omega$ significantly exceeds $\gamma_0$, the temporal LPI growth rate. This prediction has found support in various recent analytical and numerical studies\cite{zhao2017,Follett2019,zhao2019,Follett2021,Zhao2023}. In these works, it's shown that $\Delta \omega$ in the range of several percent of $\omega_0$ can effectively suppress LPI when $\gamma_0$ is on the order of $10^{-3}\omega_0$.
These studies encompass diverse broadband models, and the coherent time $\tau_c$ has been identified as a critical factor\cite{Follett2019} that determines the linear growth rates of both TPD and Stimulated Raman Scattering (SRS). In general, finite bandwidths consistently hinder the linear growth of TPD, with the extent of this effect being contingent on $\Delta\omega$.

However, when we turn to the generated hot electrons, a major concern of LPI and mainly generated during the nonlinear stage\cite{yan2012}, current numerical and experimental results have revealed qualitatively different behaviors among the broadband schemes with and without substantial intensity modulations. 
In cases involving phase modulations introduced by techniques such as SSD\cite{Skupsky1989} smoothing (with $\Delta\omega/\omega_0\leq0.1\%$) or theoretical models utilizing sinusoidal modulations of the center frequency, which lack substantial intensity modulations, prior experiments\cite{Turnbull2020b,Theobald2015a} and simulations\cite{Zhao2015,Wen2021} consistently demonstrate the suppression of LPI and a reduction in hot electron energy.
In contrast, instances with
significant intensity modulations or even prominent spikes have oddly resulted in an increased generation of hot electrons when compared to single-frequency drivers. This has been observed in experiments\cite{Wang2023} conducted at the Kunwu laser facility\cite{Ji2019,Gaoyanqi2019,JiLailin2020} with $\Delta\omega/\omega_0\sim0.55\%$ in the Shanghai Institute of Laser Plasma. In the experiments, suppression of backscatter SRS is observed coinciding with an increase in hot electron energy, which suggests that TPD could be a potential contributor to the unusual rise in hot electrons. Additionally, simulations\cite{Liu2022} with broadband high laser intensities (around $2-3\times10^{15} W/cm^2$) show enhanced hot electrons due to inflatary SRS.

It's important to note that the increased hot electron energy is achieved specifically for moderate $\Delta\omega\sim\gamma_0$, which can be common in current and anticipated broadband laser facilities. For instance, Kunwu laser facility has $\Delta\omega/\omega_0$ up to $\sim0.55\%$. Meanwhile, the FLUX\cite{dorrer2023development} laser facility under construction at the Laboratory for Laser Energetics is aiming for a fractional bandwidth of $\Delta \omega/\omega_0 > 1 \%$, and the wave field models\cite{Dorrer2020} for this facility exhibit substantial intensity modulations. 
While such a bandwidth may satisfy the condition $\Delta \omega >>\gamma_0$ for weak to moderate pumps, it might not hold true for scenarios with high laser intensities, especially in areas like laser filaments\cite{Kruer2003} and speckles\cite{Kato1984,Wen2019,Follett2022}, or during the ignition spike pulse in the Shock Ignition scheme\cite{Betti2007}. This suggests the necessity of investigating LPI responses for moderate bandwidths with $\Delta\omega\sim\gamma_0$.

These investigations usually require  precise descriptions of the waveforms of intensity-modulated broadband laser fields, which is quite challenging to obtain. Nonetheless, given that these fields always consist of numerous modulation structures differing in amplitude and duration, an initial fundamental step is to examine the impact of different amplitudes and durations of intensity modulations.
In this article, we present theoretical and numerical investigations on the TPD and hot electrons productions with the laser intensity envelope
modulated by frequency $\Delta\omega_m$. We adopt a basic two-color beam to conveniently adjust $\Delta\omega_m$ to study the TPD responses and hot electron energy for a certain range of $\Delta\omega_m$. With a series of fully-kinetic particle-in-cell (PIC) simulations scanning typical ranges of ICF laser plasma conditions for which TPD dominates near the quarter-critical density, we observe that the TPD and resulting hot electrons are driven intermittently at frequency $\Delta\omega_m$ for $\Delta\omega\sim\gamma_0$, leading to an enhancement in hot electron energy when $\Delta\omega_m$ is below a certain threshold value, denoted as $\Delta\omega_c$. We further identify the dependencies of this threshold on key laser and plasma parameters.

This paper is organized as follows. In Section II, we
describe the initial conditions of the simulations, together with the details of results and analysis. Further discussions are given in Section III. We summarize our conclusions in Section IV.

\section{\label{sec:level2}Simulations}
\subsection{Setup}
\begin{figure}[h!]
\centering
\includegraphics[width=0.42\textwidth]{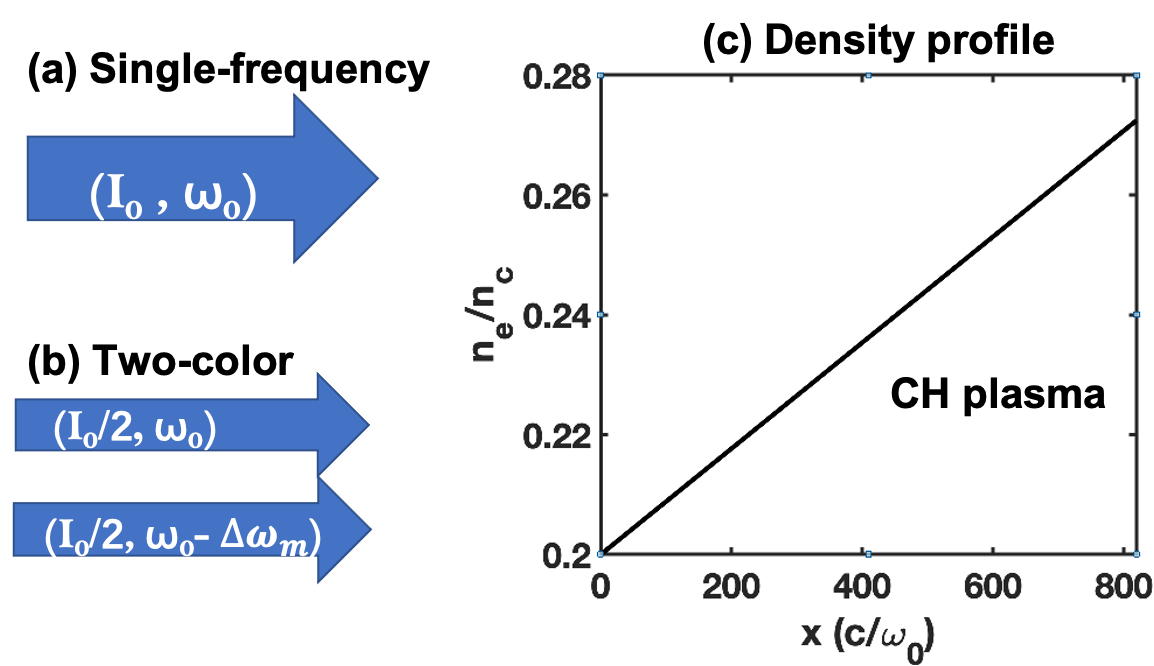}
\caption{Schematic view of the simulation setup. The left blue arrows stand for the incident laser pulses in two modes: (a) Single-frequency with intensity $I_0$ and frequency $\omega_0$; (b) Two-color with intensity $I_0$ distributed equally to two components with frequency $\omega_0$ and $\omega_0-\Delta\omega_m$. The right figure (c) demonstrates the initial density profile of the plasma.} 
\label{fig:setup}
\end{figure}

We perform a series of 2D particle-in-cell(PIC) simulations using OSIRIS\cite{Fonseca2002} code. In these simulations, the initial conditions of electron and ion temperatures $T_{e,i}$, plasma density scale lengths $L_n$ and time-averaged laser intensities $I_0$ are chosen within specific ranges of 1-3 keV, 150-300 $\mu m$ and 3-11 $\times10^{14} W/cm^2$, respectively, typical for direct-drive ICF [table (\ref{tab:simu})] with the TPD threshold\cite{Simon1983} factor $\eta$ covering 1.2 (right above threshold where $\eta=1$) to 6.7 (well above threshold). We do not adopt longer $L_n$ with higher $T_e$ and $I_0$ to isolate a TPD-dominant regime, in which the maximum SRS and SBS reflectivity among all our PIC simulations is $\sim1\%$ and $\sim3\%$, respectively. The electron density profile $n_e(x)$ of the CH plasma is linear along x-direction covering 0.2-0.27$n_c$ [figure (\ref{fig:setup})], where $n_c$ is the critical density for the laser pulses with wavelength $\lambda_0=1/3$ $\mu m$. The laser pulses are launched at the $x=0$ boundary as plane waves propagating along the x-direction and linearly-polarized along y-direction. The simulation domain size is $819.2c/\omega_0\times640c/\omega_0$ ($43.5\mu m\times 34.0\mu m$) divided evenly into $4096 \times 3200$ cells, where $c$ is the light speed in vacuum. Each cell is a $0.2c/\omega_0\times0.2c/\omega_0$ square initially distributed 200 numerical particles (100 electrons, 50 Carbon atoms and 50 protons). 
The simulations are advanced with the time step $dt=0.1414\omega_0^{-1}=0.025fs$ until $\sim8 ps$.
We adopt open electromagnetic field boundaries and thermal particle boundaries along the longitudinal direction (x-direction), and periodic boundaries for the transverse direction (y-direction). During the simulations, those electrons crossing the upper (right side) x-boundary with energy above 50keV are recorded as hot electrons.

\begin{table}
\caption{\protect\label{tab:simu} Initial simulation parameters ($L_n$, $T_e$, $T_i$, $I_0$, TPD threshold factor $\eta$) and the effective electron temperature $T_{e,eff}$ used for the $\nu_{ei}$ calculation in equation (\protect\ref{eqn:nu_ei}) and figure (\protect\ref{fig:four}).}
\begin{indented}
\item[] \begin{tabular}{ccccccccc}
\br
 Index&$L_n$ &$T_e$ &$T_i$ &$I_0$ & $\eta$  &$T_{e,eff}$\\
 &($\mu$m)&(keV)&(keV)&($\times 10^{14}W/cm^2$) & &(keV)\\
\hline
(i)& 150 & 3.0 & 1.5 &11& 2.2 &3.5\\
(ii)& 150 & 3.0 & 1.5 &6.0& 1.2 &3.5\\
(iii)& 150 & 2.0 & 1.0 &11& 3.4 &2.5\\
(iv)& 150 & 2.0 & 1.0 &6.0& 1.8 &2.4\\
(v)& 150 & 1.0 & 0.5 &11& 6.7 &1.7\\
(vi)& 300 & 3.0 & 1.5 &5.5& 2.2 &3.3\\
(vii)& 300 & 3.0 &1.5 &3.0& 1.2 & 3.1\\
\br
\end{tabular}
\end{indented}
\end{table}

To model different $\Delta\omega_m$ of the incident laser, we
employ a basic two-color scheme in which the wave field $\vec{E}_0(x,t)$ consists of two plane-wave components with equal electric field amplitude $E_1$ but different frequencies and wave nubmers of ($\omega_0$, $k_0$) and  ($\omega_0'$, $k_0'$):

\begin{align}
&\vec{E}_{0}(x,t) \nonumber \\
&=\hat{y}|E_1|[cos(k_0x-\omega_0t+\varphi_0) + cos(k_0'x-\omega_0't+\varphi_0') ]\nonumber \\
&= 2\hat{y}|E_1|cos(\frac{k_0+k_0'}{2}x-\frac{\omega_0+\omega_0'}{2}t+\frac{\varphi_0+\varphi_0'}{2}) \nonumber \\
&\quad \quad \cdot cos(\frac{k_0-k_0'}{2}x-\frac{\omega_0-\omega_0'}{2}t+\frac{\varphi_0-\varphi_0'}{2})
\label{eq:one}
\end{align}
where $\hat{y}$ refers to the unit vector along the y-direction. The $\varphi_0$ and $\varphi_0'$ are random constant phases of the two plane waves. Assuming the frequency gap $\Delta\omega_m=\omega_0-\omega_0'$ with $\Delta\omega_m<<\omega_0$, we have

\begin{align}
\vec{E}_{0}(x,t) &\approx 2\hat{y}|E_1|cos(k_0x-\omega_0t) cos[\frac{\Delta \omega_m}{2}(x/v_g-t)]
\label{eq:E0}
\end{align}
where the light group velocity $v_g$ is a slow-varying variable for our density range, and the random phases are neglected for convenience.
The above equation (\ref{eq:E0}) shows a high-frequency ($\omega_0$) oscillation  of $\vec{E}_0$ modulated by a low-frequency modulation at $\Delta\omega_m/2$. Consequently, the intensity envelope $I_m\propto E_0^2$ varies at low-frequency $\Delta \omega_m$:

\begin{align}
I_m(x,t,\Delta\omega_m)= I_0 \{1+cos[\Delta \omega_m(x/v_g-t)]\}
\label{eqn:Im}
\end{align}
where $I_0$ is the time-averaged value of $I_m(x,t,\Delta\omega_m)$ at any space location. 



To investgate TPD excited by such pump driver with $\Delta\omega_m$ at the same order with TPD growth rate $\gamma_0$, we first evaluated $\gamma_0$ using the formula\cite{Kruer2003}
\begin{align}
\label{eqn:gamma0}
\gamma_0\approx \frac{k_yv_{os}}{4}|\frac{(k_x-k_0)^2-k^2}{k(k_x-k_0)}|
\end{align}
where $k_x$ and $k_y$ are the $x$ and $y$ components of $k$, the wave number of the TPD daughter plasma waves. The pump intensity $I$ is associated with $v_{os}$ via $I\propto v_{os}^2$, the oscillation velocity of electrons in the wave fields. Since the maximum $\gamma_0$ at certain $n_e$ always stays on the TPD hyperbola\cite{Kruer2003} $k_y^2=k_x(k_x-k_0)$, we find $1\times10^{-3}<\gamma_0/\omega_0<2\times10^{-3}$ holds for all the parameters in table (\ref{tab:simu}). Therefore, we scan $\Delta\omega_m/\omega_0$ from $0.05\%$ to $1.0\%$, covering both $\Delta\omega_m\leq\gamma_0$ and $\Delta\omega_m\gg\gamma_0$.

\subsection{Results: Anomalous hot electron energy for two-color drivers}

\begin{figure}
\includegraphics{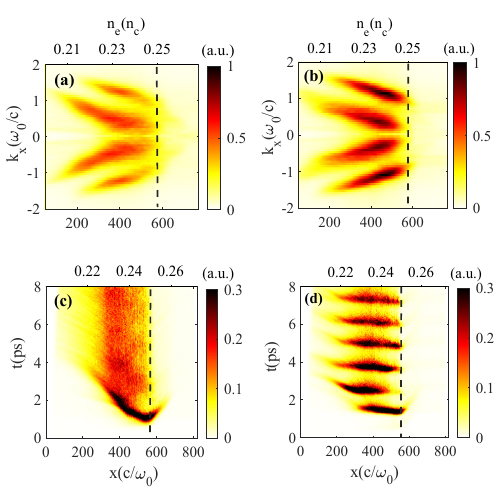}
\caption{ PIC simulation results for the case (i) of table (\ref{tab:simu}). Typical $k_x-x$ spectra of $E_x$ fields are plotted (at $t\sim3.7ps$) in (a)(b) using the moving-window method along x-direction\cite{yan2012}, and the space-time evolutions of $<E^2_x>$ are shown in (c)(d). Among the four figures, (a)(c) and (b)(d) are respectively for single-frequency driver and  two-color driver in $k_x-x$ with $\Delta\omega_m/\omega_0=0.1\%$. The vertical dashed lines mark the location of $n_c/4$ surfaces. 
\protect\label{fig:Ex-x-t}}
\end{figure}

We perform a series of PIC simulations with constant $I_0$ and varying $I_m(x,t,\Delta\omega_m)$ [equation (\ref{eqn:Im}) with varying $\Delta\omega_m$] for each parameter set in table (\ref{tab:simu}). 
In all the simulations, TPD is observed to be the dominant LPI and source of hot electrons. 
For both the two-color and single-frenquency drivers, TPD modes are evident with similar characteristic quadruple-branch structure\cite{yan2012} near $n_c/4$ surface in $k_x-x$ plots of $E_x$ (the electric field along x-direction) [figure \ref{fig:Ex-x-t}(a)(b)], indicating that most $E_x$ fields are from the EPWs of TPD modes. 


Despite similar space spectra, the temporal behaviors of TPD are different with and without the intensity modulation introduced by the frequency shift $\Delta\omega_m$. Figure \ref{fig:Ex-x-t}(c)(d) show the space and time evolutions of $<E_x^2>$ for single-frequency laser pulse and two-color with $\Delta\omega_m/\omega_0=0.1\%$, 
where $<E_x^2>$ is the $E_x$ field energy which directly correlates with TPD modes energy, and the brackets in $<E_x^2>$ stand for the space average over y-direction. 
For single-frequency laser pulse [figure \ref{fig:Ex-x-t}(c)], the driven TPD modes first grow near the $n_c/4$ density, then diffuse to the lower density region until reaching a quasi-steady state, the same as the previous research with similar physical conditions \cite{yan2012}.
However, when the laser intensity is modulated by $\Delta\omega_m$, the excited TPD modes exhibit intermittent character with period of $\sim1.1ps\approx2\pi/\Delta\omega_m$ [figure \ref{fig:Ex-x-t} (d)], showing that 
TPD modes undergo intermittent cycles of growth and decay, synchronized with the periodic variations in laser intensity.

\begin{figure}
\includegraphics{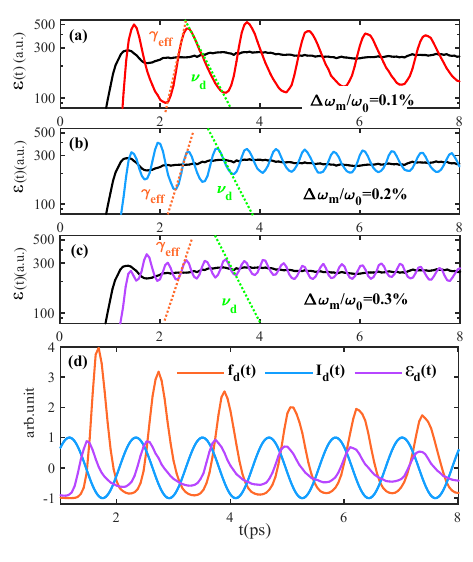}
\caption{\protect\label{fig:two} PIC simulation results for the case (i) of table (\ref{tab:simu}). The time evolutions of the integrated longitudinal electric field energy $\epsilon(t)$ for (a) $\Delta\omega_m/\omega_0=0.1\%$(red);(b) $\Delta\omega_m/\omega_0=0.2\%$(blue);(c) $\Delta\omega_m/\omega_0=0.3\%$(purple). In (a), (b) and (c), the black lines correspond to single-frequency laser pulse with the same average intensity $I_0$, and the slopes of the dotted orange and green lines stand for the effective growth rate $\gamma_{eff}$ and decay rate $\nu_d$ of $\epsilon(t)$ in each oscillation period. (d) The normalized deviations of laser intensity $I_d(t)$(blue), $E_x$ field energy $\epsilon_d(t)$(purple) and hot electron energy $f_d(t)$(orange) defined by equation (\ref{eqn:Id}-\ref{eqn:fd}) for two-color laser pluse with $\Delta\omega_m/\omega_0=0.1\%$. 
}
\end{figure}

Along with the intermittency, the TPD modes energy $\varepsilon(t)=\int{<E_x^2>dx}$ oscillates around $\varepsilon_{I0}$, the steady-state value of $\varepsilon(t)$ for the simulation with constant $I_0$. 
Since the peak intensity for the two-color laser pulses is $2I_0$ [equation (\ref{eqn:Im})], the TPD modes driven at the peak intensity should be stronger than $\varepsilon_{I0}$, as the modulation period is sufficient for the doubled power to take effect. Similarly, $\varepsilon(t)$ should become lower than $\varepsilon_{I0}$ for the valley laser intensity. 
All these oscillations are demonstrated in figure (\ref{fig:two}), which compares the strength of TPD modes driven by constant $I_0$ and varying $I_m(x,t,\Delta\omega_m)$ in (a)-(c) for 
$\Delta\omega_m/\omega_0$ of $0.1\%$, $0.2\%$ and $0.3\%$, respectively. 
Particularly, the oscillation amplitude increases for smaller $\Delta\omega_m$, as it causes longer growing and decaying time for TPD modes energy to reach higher peaks and lower valleys.

\begin{figure}
\includegraphics{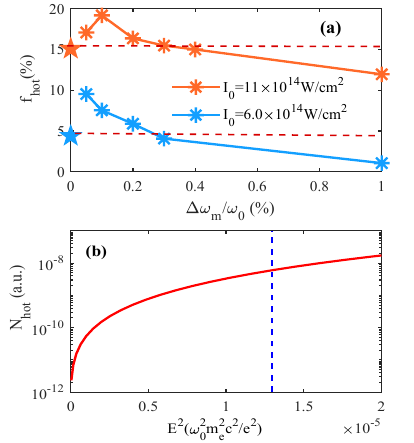}
\caption{\protect\label{fig:three} (a) Time-averaged hot electron energy fraction of the incident two-color laser pulse, $f_{hot}$, for different $\Delta\omega_m$ and average intensities $I_0=6\times10^{14}W/cm^2$ (blue) and $11\times10^{14}W/cm^2$ (orange). 
The two pentagram markers at $\Delta\omega_m=0$, in conjunction with the horizontal red dashed lines, indicate the steady-state $f_0$ values for single-frequency drivers.
(b) The dependence of $N_{hot}$ on the energy of EPW eletric field energy obtained from equation (\ref{eqn:Nhot}). The maximum $E^2$ among all the PIC simulations are marked by the vertical blue dashed line.}
\end{figure}

By modulating the oscillation ampitudes and periods of the TPD modes energy, $\Delta\omega_m$ can affect the energy of hot electrons, since the TPD modes are the primary source of hot electron generation.
This is illustrated in figure (\ref{fig:three}a) by the results of PIC simulations with the setup (i) and (ii) of table (\ref{tab:simu}). 
We find that the ratio of the time-averaged hot electron energy to the incident laser energy during the quasi-steady state, $f_{hot}$, changes significantly with $\Delta\omega_m$. 
At the limit of large $\Delta\omega_m$, $f_{hot}$ drops with increasing $\Delta\omega_m$ while staying well below the hot electron energy ratio $f_0$ for constant $I_0$, agreeing with the trend inferred by the previous study\cite{thomson1974} for $\Delta\omega>>\gamma_0$. 
However, when $\Delta\omega_m$ is smaller than a threshold frequency $\Delta\omega_c$, we suprisingly find that $f_{hot}$ goes beyond $f_0$.
In other words, 
when contrasted with the ideal TPD matching condition associated with a single-frequency driver, the two-color drivers with compromised matching conditions can anomalously produce significantly greater amounts of hot electron energy. This observation implies the existence of unidentified mechanisms contributing to the augmentation of $f_{hot}$.

To grasp the mechanism responsible for the anomalous $f_{hot}$, we explore the temporal evolutions of laser intensity
$I_m(x=0,t,\Delta\omega_m)$, total electric field energy $\varepsilon(t)$ of TPD modes and the instantaneous hot electron energy ratio $f_{\Delta\omega_m}(t)$.  For $\Delta \omega_m$ corresponding to $f_{hot}>f_0$, all these quantities oscillate around the stead-state values in the constant $I_0$ case 
[equation (\ref{eqn:Im}) and figure (\ref{fig:two}a-c)] after the initial linearly growing stage ($\sim 1ps$), so we turn to their normalized deviations from the steady-state values for constant $I_0$ case: 

\begin{align}
\label{eqn:Id}
I_d(t)&=[I_m(t)-I_0]/I_0 \\
\label{eqn:epsilond}
\varepsilon_d(t)&=[\varepsilon(t)-\varepsilon_{I0}]/\varepsilon_{I0} \\
\label{eqn:fd}
f_d(t)&=[f_{\Delta\omega_m}(t)-f_0]/f_0
\end{align}
where the arguments $x$ and $\Delta\omega_m$ in $I_m(x=0,t,\Delta\omega_m)$ are neglected for convenience, and the subscript $d$ and $0$ denote the normalized deviation and the corresponding steady-state values, respectively
. The time evolutions of these quantities are plotted in figure (\ref{fig:two}d) for the simulation with $\Delta\omega_m/\omega_0=0.1\%$, in which the overall $f_{hot}$ exceeds $f_0$ notably (figure \ref{fig:three}a). 
We find the laser intensity envelope $I_d(t)$ oscillates between 0 and 2, consistent with equation (\ref{eqn:Im}). 
As expected, $\varepsilon_d(t)$ and the hot electron energy $f_d(t)$ also oscillate following $I_m(t)$ with visible delays caused by the growth time of TPD and travel time for hot electron to reach the boundary. 
Remarkably, $f_d(t)$ has much higher peaks than $\varepsilon_d(t)$ and $I_d(t)$ by average factors of 3.3 and 2.3, respectively. 
This reveals the direct cause of the anomalous $f_{hot}$: 
The instantaneous hot electron energy can experience significantly greater increases compared to the increase in laser intensity.
Particularly, when the hot electron energy ratio during peak periods of $I_d(t)>1$ surpasses $2f_0$, the overall $f_{hot}$ undoubtedly surpasses $f_0$,
regardless of the amount of hot electrons generated during the intensity valleys ($I_d(t)>1$).



The over increase of $f_d(t)$ should be relevant to the scaling between $\varepsilon(t)$ and $f_{hot}$. 
Previous studies have shown that hot electrons are usually accelerated through complex staged processes\cite{yan2012} or cavitating Langmuir turbulence\cite{vu2012}, but the hot electron generation efficiency can still be reasonably evaluated by calculating the number of thermal electrons that can be caught and accelerated by the EPW field potentials.
Generally, EPWs can catch and accelerate the electrons with velocities around the EPW phase velocities
$v_p$ from $v_p-\Delta v$ to $v_p+\Delta v$ , with $\Delta v$ depending on the plasma wave potential
$\Phi$ 
\begin{equation}
\Delta v=\sqrt{\frac{2e\Phi}{m_e}}=\sqrt{\frac{eE}{km_e}}
\end{equation}
where $k$ and $E=k\Phi/2$ are the wave number and field amplitude of the EPW. For a Maxwellian velocity distribution $f_M$, the amount of electrons that can be caught and accelerated by EPWs are 

\begin{equation}
\label{eqn:Nhot}
N_{hot}=\int_{v_p-\Delta v}^{v_p+\Delta v}f_M dv
\end{equation}

\textcolor{black}{For forward-going EPW modes with wave number $k$ from $0.8$ to $2 k_0$ and plasma $T_e$ from 1 to 3keV }
, we calculate $N_{hot}$ with equation (\ref{eqn:Nhot}) for different wave fields
. Representative results are plotted in figure (\ref{fig:three}b) for $k=1.3k_0$ and $T_e=3keV$, showing that $N_{hot}$ approximately increases exponentially as $E^2$ rises, implying that a slight increase in the EPW amplitude can effectively capture and accelerate a significantly larger number of thermal electrons. This phenomenon offers a qualitative explanation for the high oscillating peaks observed in $f_{d}(t)$ [figure~\ref{fig:two}(d)] and the condition $f_{hot}>f_0$ (figure \ref{fig:three}a).
For increasing $\Delta\omega_m$, the growth time for TPD modes decreases and causes lower $\varepsilon(t)$. According to figure (\ref{fig:three}b), this causes lower $f_{hot}$, qualitatively agreeing with the trend in figure (\ref{fig:three}a).

\subsection{Modeling the $f_{hot}$ dependence on laser plasma conditions}

Now we investigate the $f_{hot}$ dependence on $\Delta\omega_m$ and other core quantities, with particular focus on characterizing the transition region near $f_{hot}=f_0$. The hot electron generation, as mentioned previously, occurs mainly in TPD nonlinear stage associated with complex staged-acceleration\cite{yan2012} and Langmuir turbulence\cite{vu2012}. Thus, it is unpractical to accurately calculate $f_{hot}$ considering all the details. However, with the most essential factors properly distilled, we are able to construct a simplified scaling model that provides an estimation of $\Delta\omega_m$ near the transition region $f_{hot}\sim f_0$ as well as the behavior in the vicinity. 

Previous discussions have shown the sensitivity of $f_{hot}$ to the magnitude of the EPW field energy $\varepsilon(t)$, so the model describing $f_{hot}$ should prioritize the amplitude of oscillating $\epsilon(t)$ and the time duration for $\epsilon(t)>\epsilon_{I0}$.
As $\epsilon(t)$ oscillates around $\epsilon_{I0}$ at frequency $\Delta\omega_m$, the time duration of $\epsilon(t)>\epsilon_{I0}$ can be approximated by $\pi/\Delta\omega_m$, half of the oscillation period $2\pi/\Delta\omega_m$.
During the period, the amplitude of $\epsilon(t)$ depends on 
the effective growth and decay rates, $\gamma_{eff}$ and $\nu_{d}$ respectively [figure (\ref{fig:two}a-c)], and associated durations $\tau_{\gamma}$ and $\tau_{\nu}$, which are defined as the time intervals between neighboring peaks and valleys. 
The sum of $\tau_{\gamma}$ and $\tau_{\nu}$ equals a single period 
$2\pi/\Delta\omega_m$, and 
the precise allocation of time is determined by the interplay of the dynamic growth and decay mechanisms. Despite the complexity, higher $\epsilon(t)$ amplitude should favor higher $\gamma_{eff}$ and lower $\nu_{d}$. 
In essence, the leading determinants of $f_{hot}$ are $\gamma_{eff}$, $\nu_d$ and $\Delta\omega_m$. 

While $\Delta\omega_m$ is clear as the frequency of the intensity modulation envelope, the calculations of both  $\gamma_{eff}$ and $\nu_d$ are vague. For a constant laser intensity and zero damping condition, the existing theory\cite{Rosenbluth1973,Kruer2003} shows that the TPD growth rate is proportional to the square root of the laser intensity (equation \ref{eqn:gamma0}). However, the growth rate for dynamic laser intensity and finite damping rate is not well understood yet. Besides, the decay of TPD energy is governed by multiple mechanisms including collisional damping, Landau damping, mode transition, etc. Some of these mechanisms depend on modes locations in space and their overall effects on $\varepsilon(t)$ are challenging to calculate in a standard way. For example, Landau damping is negligible near $n_c/4$ surface, but significant at lower density regions with $k\lambda_D>0.25$, where $\lambda_D$ is the Debye length of the background plasma; mode transition occurs mainly near the reflection point of the EPWs near $n_c/4$ surface. Moreover, the laser intensity during the $\varepsilon(t)$ decay stage is non-zero and the contribution of the laser pump to the decaying rate is still unknown. The nonlinear processes like secondary instabilities can also affect both growth and decay processes. All these complications cause extreme difficulties in obtaining an accurate theoretical form of the effective decay rate.

We adopt the following approximations to grasp the essential factors. First, since the $\gamma_{eff}$ is powered by the incident laser, we choose $2\gamma_0-\nu_d$ to scale the variations of $\gamma_{eff}$, where $\gamma_0$ is the amplitude (not energy) growth rate of TPD EPWs and calculated by equation (\ref{eqn:gamma0}) for $I=I_0$. Second, $\nu_d$ only includes the collisional damping with other physics processes in the decaying stage neglected. So we have $\nu_d\approx \nu_{ei}$, the EPW energy damping rate caused by electron-ion collisions \cite{Kruer2003}
\begin{align}
\label{eqn:nu_ei}
\nu_{ei}=3\times10^{-6}ln\Lambda\frac{n_eZ}{T_e^{3/2}}
\end{align}
in which $ln\Lambda$ is the Coulomb logarithm\cite{Huba2013}, $n_e$ is the electron density in unit of $cm^{-3}$, $Z=5.3$ is the average charge state of CH ions, and $T_e$ is the electron temperature in unit of $eV$.

The approximation of using $I=I_0$ and associated $\gamma_0$ to estimate the growth power should be valid, as $I_0$ represents the average laser intensity. However, the approximation of the damping process appears overly bold as it overlooks numerous unassessed processes which can either enhance or slow down the damping. This oversight is likely to lead to a significant deviation of the decaying rate of TPD energy.
In fact, the effective values of $\nu_d$ obtained from fitting simulation data deviate from $\nu_{ei}$ by up to $\pm50\%$.
However, the approximation $\nu_d\approx\nu_{ei}$ still captures the overall variation of $\nu_d$. For larger $\nu_{ei}$, $\nu_d$ trends greater, and \textit{vice versa}.
Therefore, although the omission of various decaying-relevant effects may affect the comprehensive quantitative calculation (a step we do not taken in this manuscript), it might not have a substantial impact on our scaling analysis.

Using these approximations, we formulate a dimensionless factor $H$ that characterizes the dependency of $f_{hot}$ on $\Delta\omega_m$, $\gamma_0$ and $\nu_{ei}$:
 \begin{equation}
 \label{eqn:H}
 H=\frac{\Delta\omega_m}{(\gamma_{eff}-\nu_{d})/2}=\frac{\Delta\omega_m}{\gamma_{0}-\nu_{ei}}
 \end{equation}
in which the factor $1/2$ in the denominator translates the energy growth and decay rates to amplitude rates. This formula clearly show that $H$ increases as $\Delta\omega_m$ rises, or $\gamma_0$ and $\nu_{ei}$ decreases, opposite to the trend of $f_{hot}$. Particularly, when $H\sim1$, we have $\Delta\omega_m\sim\gamma_0$, correspoinding to moderate bandwidth that can cause $f_{hot}\sim f_0$.

\begin{figure}
\includegraphics{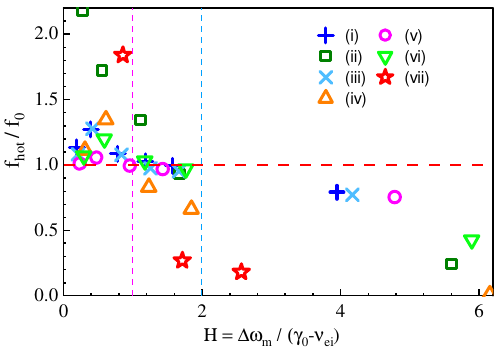}
\caption{\protect\label{fig:four} The PIC simulation results of hot electron energy ratio $f_{hot}/f_0$ for different $H$ defined by equation (\ref{eqn:H}).
The correspondences between marker styles and simulation parameter sets of table (\ref{tab:simu}) are illustrated near the top-right corner. For the simulation set (i), the six data points correspond to $\Delta\omega_m/\omega_0=0.05\%$, $0.1\%$, $0.2\%$, $0.3\%$, $0.4\%$ and $1\%$ from left to right, respectively. All other simulation sets have five data points with $\Delta\omega_m/\omega_0$ of $0.05\%$, $0.1\%$, $0.2\%$, $0.3\%$ and $1\%$.
\textcolor{black}{Please note that the leftest ($\Delta\omega_m/\omega_0=0.05\%$, $H=0.43$ and $f_{hot}/f_0=3.31$) and rightest ($\Delta\omega_m/\omega_0=1\%$, $H=8.58$ and $f_{hot}/f_0=0$ ) data points of the simulation set (vii) are not displayed as they are outside the range of this figure}.
}
\end{figure}

The PIC simulation results of $f_{hot}/f_0$ and corresponding $H$ factors are plotted in figure (\ref{fig:four}). 
Across varied physical conditions with TPD threshold factor $\eta$ spanning from 1.2 to 6.7, the simulations consistently reveal lower $f_{hot}$ for higher H, except in the quite low $H$ range for certain simulations (discussed in section \ref{sec:low_omega}). This aligns well with our analysis of the correlation between $H$ and $f_{hot}$.
Furthermore, for all the parameter sets, the transition point $f_{hot}/f_0=1$ consistently falls in $1<H<2$, indicating the validity of the $H$ defined in the preceding equation (\ref{eqn:H}).


To provide clarity, we elaborate the computation specifics of $\nu_{ei}$ in equation (\ref{eqn:H}). The challenge lies in selecting the appropriate $T_e$. While the initial $T_e$ is provided in table (\ref{tab:simu}), it appreciably increases during the 8ps simulations due to collisional heating. Therefore, the suitable $T_e$ for $\nu_{ei}$ should not be the initial value but rather the average value over the time duration used for calculating $f_{hot}$ 
, which is from approximately $t\sim1.5 ps$ to $\sim8ps$, with the starting and ending times appropriately adjusted to cover full oscillation periods of $f_d(t)$.
The actual effective $T_e$ values are listed in table (\ref{tab:simu}) as $T_{e,eff}$.

 
\subsection{The $\Delta\omega_m$ threshold for $f_{hot}=f_0$}

\begin{figure}
\includegraphics{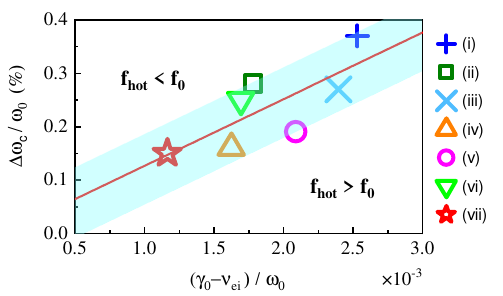}
\caption{\protect\label{fig:five} The threshold $\Delta\omega_c$  for $\gamma_0-\nu_{ei}$ of different initial laser plasma conditions. The seven markers correspond to seven simulations in table (\ref{tab:simu}). The dark red straight line denotes the fitted results of the seven dots with the surrounding light blue region covering vertical variation $\pm 0.05\%$. The upper-left and lower-right white areas stand for the parameters corresponding to $f_{hot}<f_0$ and $f_{hot}>f_0$, respectively.}
\end{figure}

The intensity modulation frequency $\Delta\omega_m=\Delta\omega_c$ associated with $f_{hot}=f_0$ is of significant practical importance, as it provides references to evaluate the hot electron energy change for certain broadband LPI experiments, given the intensity modulation descriptions of the laser pulses. 
Using the defined $H$ in equation (\ref{eqn:H}) and the simulation results in figure (\ref{fig:four}), we obtain a $H$ value for each simulation parameter set through linear interpolation of the simulation data. This $H$ value tells the threshold $\Delta\omega_c$ for different laser and plasma conditions with respective $\gamma_0-\nu_{ei}$. 
The relation between $\Delta\omega_c$ and $\gamma_0-\nu_{ei}$ is given in figure (\ref{fig:five}), which shows an approximate linear scaling 
\begin{align}
\label{eqn:omegaC}
\Delta\omega_c\sim 1.25(\gamma_0-\nu_{ei})
\end{align}
with all data located within $\pm 0.05\%\omega_0$. 
For given laser intensities and plasma conditions, $\gamma_0$ and $\nu_{ei}$ can be calculated with equation (\ref{eqn:gamma0}) and (\ref{eqn:nu_ei}). Then the $\Delta\omega_c$ can be estimated with the scaling equaiton (\ref{eqn:omegaC}). 
When $\Delta\omega_m$ exceeds $\Delta\omega_c$ significantly into the upper left area in Fig.~\ref{fig:four}, TPD is suppressed resulting in fewer hot electrons compared to single-frenquency driver. Conversely, when $\Delta\omega_m$ falls within the lower right area (where $\Delta\omega_m$ is obviously smaller than $\Delta\omega_c$), enhencements in hot electron production occur due to intensity modulation. 

\section{Discussions}


\subsection{Considering realistic wave fields of broadband laser pulses}
While equation (\ref{eqn:omegaC}) clearly demonstrates a scaling relationship for the modulation frequency threshold $\Delta\omega_c$ with laser plasma conditions, it cannot be directly applied to the waveforms of realistic broadband laser pulses, which typically consist of numerous intensity modulations, each with different amplitudes and durations. Moreover, the actual shape of these modulations rarely resembles a simple sinusoidal curve oscillating from zero to 2$I_0$.
However, it's important to note that the core physics governing the enhancement of $f_{hot}$ doesn't hinge on the precise shape of the modulation. It merely requires the presence of substantial modulation peaks lasting approximately as long as a specific duration on the order of $1/\gamma_0$, and such modulations should be present in realistic broadband laser pulses with moderate bandwidths. First, intensity modulations are common in the modeling of contemporary and future broadband laser fields\cite{Ji2019,Gaoyanqi2019,JiLailin2020,Dorrer2020}, where intensities fluctuate across a range from zero to several times $I_0$ with a distribution that depends on the laser amplification process. Second, the durations of the intensity peaks are influenced by the actual bandwidth $\Delta\omega$ of the laser pulses. 
Higher $\Delta\omega>>\gamma_0$ corresponds to shorter duration. But since the duration should increase with $\Delta\omega$, we always have the duration comparable to $1/\gamma_0$ for a certain range of moderate $\Delta\omega$.


In contrast to equation (13), which was derived using a sinusoidal intensity modulation as a basis, we can propose a more general expression for the bandwidth threshold of realistic intensity-modulated laser pulses:
\begin{align}
\label{eqn:omega_th}
\Delta\omega_{th} \propto\gamma_0-\nu_{ei} 
\end{align}
where $\Delta\omega_{th}$ is the threshold of the bandwith $\Delta\omega$, which should be proportional to the effective intensity modulation frequency $\Delta\omega_{m,eff}$ for certain broadband schemes. 
The precise connection between $\Delta\omega_{th}$ and $\gamma_0-\nu_{ei}$ depends on the relation of $\Delta\omega$ and $\Delta\omega_{m,eff}$, which relies on the wave field modeling for specific broadband scenarios. For broadband laser facilities, this still needs further considerable investigation. But for certain cases which tune frequencies\cite{Michel2009} of different laser beams to form a finite bandwidth, the waveforms should be relatively more convenient to obtain. Nevertheless, this topic falls beyond the current scope of this paper and will be investigated in future research.

\subsection{$f_{hot}$ at low $\Delta\omega_m$ limit}
\label{sec:low_omega}
Given that $\Delta\omega_m<\Delta\omega_c$ anomalously results in a higher value of $f_{hot}$, the question arises whether $f_{hot}$ exhibits a monotonic increase for smaller $\Delta\omega_m$. This question can be investigated by examining the scenario at the lower $\Delta\omega_m$ limit, representing intensity modulations with a long period of $2\pi/\Delta\omega_m$. This implies alternations between intensity peaks at approximately $2I_0$ and valleys at nearly $0$ intensity, each with a duration of $\pi/\Delta\omega_m$. Since the valleys are expected to make negligible contributions to $f_{hot}$, the original question becomes whether a sufficiently long pulse with an intensity of $2I_0$ can generate a $f_{hot}$ value that is twice that produced by an intensity of $I_0$.

As the pulse duration is sufficient for TPD to reach quasi-steady states, we can deduce an answer to this query from prior studies\cite{Froula2012,Turnbull2020b}. These studies indicate that $f_{hot}$ experiences a rapid exponential increase with TPD parameter $\eta$, which is proportional to $I_0$, when $\eta\sim1$. This suggests that hot electron energy can increase significantly more than twofold when the intensity is doubled, enhancing overall $f_{hot}$. However, for higher $\eta\geq2$, the rate of increase in $f_{hot}$ noticeably decelerates with $\eta$, and it is more likely that $f_{hot}$ increases by less than a factor of two when the intensity is doubled, reducing overall $f_{hot}$.

These analysis align with the findings illustrated in figure (\ref{fig:four}), which shows that in most simulations with $\eta\geq1.8$, $f_{hot}$ decreases at the lowest $H$ ($\Delta\omega_m/\omega_0=0.05\%$). Two exceptions to this trend are simulations (ii) and (vii) in Table (\ref{tab:simu}) with the lowest $\eta=1.2$.

\subsection{$\Delta\omega_c$ dependences}

While equation (\ref{eqn:omegaC}) outlines the relationship between $\Delta\omega_c$ and both pump power and plasma conditions, we notice that it does not include $L_n$, a critical factor for LPI in various experiments. Here we point out that equation (\ref{eqn:omegaC}) actually describes $f_{hot}/f_0$ instead of $f_{hot}$ itself. Since the effects of $L_n$ on $f_{hot}$ with two-color laser pulses also work on $f_0$, it is not surprising to have $f_{hot}/f_0$ and $\Delta\omega_c$ weakly or not dependent on $L_n$.

This mechanism also can be relevant to the weak dependence of $\Delta\omega_c$ on $T_e$, which is only implicitly involved in $\Delta\omega_c$ through $\nu_{ei}$.
However, it's essential to acknowledge that the simulation results in figure (\ref{fig:five}) show visible patterns that
the $\Delta\omega_c$ data points for higher $T_e$ consistently exhibit slightly higher values for close $\gamma_0-\nu_{ei}$. 
This observation suggests potential existence of unaccounted factors that could enhance the precision of the $\Delta\omega_c$
scaling model outlined in equations (\ref{eqn:omegaC}) and (\ref{eqn:omega_th}). These factors will be a subject of investigation in future research endeavors.

\section{summary}
In summary, we have investigated the hot electron generation from TPD instability driven by broadband laser pulses with intensity modulations, which is common in the theoretical waveforms of current and future broadband laser facilities. 
Using a basic two-color scheme in 2D PIC simulations, our investigation has unveiled that TPD modes energy and resulting hot electron energy ratio $f_{hot}$ oscillate at the modulation frequency $\Delta\omega_m$ when $\Delta\omega_m$ is at the same order as the TPD growth rate $\gamma_0$.
If $\Delta\omega_m$ falls below a threshold frequency $\Delta\omega_c$, such oscillation results in overall $f_{hot}$ anomalously exceeding the hot electron energy ratio $f_0$ for a single-color pump.
We also characterize the dependence of $\Delta\omega_c$ and laser plasma conditions, and we provide a scaling relation to evaluate $f_{hot}/f_0$ for different laser bandwidth $\Delta\omega$.

To bridge the gap between our investigations
and the intricate waveforms in reality, the actual wave fields needs to be properly modeled to characterize the relation between the bandwidth $\Delta\omega$ and the effective modulation frequency $\Delta\omega_{m,eff}$. 
While this connection can be relatively straightforward for multi-color pumps based on frequency detuning, it can be much more complicated for existing and future broadband laser facilities. Investigations in this regard will constitute future research endeavors.

\ack{
This research was supported by the National Natural Science Foundation of China (NSFC) (Grant Nos. 12275269, 12175229, 12388101, 12275032 and 12075227), by the Strategic Priority Research Program of Chinese Academy of Sciences (Grant Nos. XDA25010200 and XDA25050400). The numerical simulations in this paper were conducted on Hefei advanced computing center. We thank Dr. Xiaohui Zhao for the insightful discussions on the wave fields of the laser pulses on Kunwu facility.}

\section*{References}
\bibliographystyle{iopart-num}
\bibliography{iopart-num}

\end{document}